\def\H{{\cal H}}
\def\R{{\bf R}}
\def\adag{{a^\dagger}}
\def\h{{\frac{1}{2}}}
\def\rh{{\frac{1}{\sqrt{2}}}}

\def\dx2{{{\rm d^2}\over{{\rm d}x^2}}}

\def\osp{{osp(1,2)}}
\def\bdag{{b^\dagger}}
\def\str{{\rm str}}
\def\res{{\rm res}}
\def\vac{{|0\rangle}}

\def\fdag{{f^\dagger}}

\def\q{{\frac{1}{4}}}
\def\sgn{{\rm sgn}}
\documentstyle[12pt]{article}
\begin{document}
\begin{flushright}
DAMTP/93-60\\
hep-th/9312099
\end{flushright}
\vskip 0.5in
{\begin{center}
\baselineskip 22pt
{\Large \bf On Integrable Models Related to the $osp(1,2)$ Gaudin
Algebra}
\vskip 0.5in
\baselineskip 13pt
T. Brzezi\'nski  and A.J. Macfarlane
 \\[.3in]
{\em  Department of Applied Mathematics \& Theoretical
Physics \\  University of Cambridge, Cambridge  CB3 9EW, U.K.}
\vspace{18pt}

December 1993
\end{center}
\vspace{18pt}
\baselineskip 14pt

\begin{quote}ABSTRACT We define the $osp(1,2)$ Gaudin algebra  and
consider integrable
models described by it. The models include the $\osp$ Gaudin magnet and
the Dicke model related to it. Detailed discussion of the simplest
cases of these models is presented. The effect of the presence of
fermions on the separation of variables is indicated.

\end{quote}

\vspace{18pt} }
\baselineskip 20pt

\noindent {\bf I. INTRODUCTION}\vskip .3cm

\noindent In 1976 \cite{gaudin1}(see also \cite{gaudin2}) M. Gaudin
introduced a quantum mechanical model of $N$ particles
with $su(2)$ or $su(1,1)$ spin, ever since known as a Gaudin model.
The model is
described by a set of $N$ hamiltonians
\begin{equation}
H\sb i = \sum\sb{j\neq i}{{\bf S}\sp i\cdot{\bf S}\sp
j\over{\epsilon\sb i - \epsilon\sb j}}, \quad i= 1,\ldots ,N,
\label{suh}
\end{equation}
where $\epsilon\sb 1 >\epsilon\sb 2 >\ldots >\epsilon\sb N$ are free
parameters and ${\bf S}\sp i$ is the $su(2)$ or
$su(1,1)$ spin of the $i$-th particle, i.e.
\begin{equation}
[S\sp i\sb z, S\sp j\sb \pm ]\sb - = \pm\delta\sb{ij}S\sp i\sb \pm ,
\quad [S\sp i\sb +, S\sp j\sb - ]\sb - = 2\delta\sb{ij}S\sp i\sb z .
\label{su2}
\end{equation}
Here $[\; ,\; ]\sb -$ denotes the commutator. The total spin ${\bf S}^2
= \sum\sb i ({\bf S}^i)^2$ and $S\sb z = \sum\sb i S\sb z\sp i$ are
remaining constants of motion. Gaudin showed that the
hamiltonians (\ref{suh}) commute with each other and constructed
their common eigenvectors by the coordinate Bethe Ansatz.  In
\cite{jurco4} the Gaudin model was generalised to any semisimple
Lie algebra.

Because of its simplicity, the Gaudin model has been used as a testing
ground for the ideas such as the functional Bethe Ansatz and the
general procedure of
separation of variables \cite{sklyanin4}. Its connection to the
Neumann model \cite{neumann1}, revealed in \cite{kuznetsov1} allows one
to generalise separation of variables for some types of coordinates on
Riemannian manifolds.  It has also applications in quantum
optics \cite{jurco3}.

In \cite{sklyanin4} the Gaudin model was solved in the framework of the
algebraic Bethe Ansatz. It was also shown there \cite{sklyanin4} that
the model is governed by a Yang-Baxter algebra, called a Gaudin
algebra, with commutation relations linear in the generators and
determined by a classical
$r$-matrix. It is to be stressed these features are present in the
model despite its quantum mechanical nature. In fact, the Gaudin model
is one of a large class of models, with such an algebraic nature, so that
study of Gaudin algebras becomes an important issue.

To construct a Gaudin algebra we need two ingredients: the
classical $r$-matrix and quantum $L$-operator. By the former we mean
an $M^2\times M^2$ matrix, of which the components are functions
satisfying the classical Yang-Baxter equation
\begin{equation}
[r\sb{12}(\lambda-\mu), r\sb{13}(\lambda-\nu)]\sb- +
[r\sb{12}(\lambda-\mu), r\sb{23}(\mu-\nu)]\sb- +
[r\sb{13}(\lambda-\nu), r\sb{23}(\mu-\nu)]\sb- = 0.
\label{class.yb}
\end{equation}
The subscripts in (\ref{class.yb}) indicate the vector spaces in
$\R^M\otimes\R^M\otimes\R^M$ on which $r$ acts non-trivially, and
$\lambda,\mu,\nu$ are spectral parameters. The quantum $L$-operator is
an $M\times
M$ matrix, of which the components are operator valued functions, such
that
\begin{equation}
[L\sb 1(\lambda), L\sb 2(\mu)]\sb - = -[r(\lambda-\mu), L\sb 1(\lambda)
+L\sb 2(\mu)]\sb -,
\label{gaudin.compact}
\end{equation}
where $L\sb 1(\lambda) = L(\lambda)\otimes 1$ and $L\sb 2(\mu) =
1\otimes L(\mu)$. The infinite dimensional algebra generated by the
components of $L(\lambda)$ is called a Gaudin algebra. If a Gaudin
algebra describes the Gaudin system corresponding to Lie algebra $\cal
G$, then we term this kind of a Gaudin algebra a $\cal G$ Gaudin
algebra. To any simple  Lie algebra one can associate different types of
Gaudin algebras, corresponding to different types of $r$-matrices
\cite{belavin2}.
For instance we can consider rational, trigonometric or elliptic Gaudin
algebras. In what follows we restrict ourselves to the rational case.
Using language just introduced  we can say that the model (\ref{suh}) is
described by the rational $su(2)$ or $su(1,1)$ Gaudin algebra.

It is easy to extend the notion of a Gaudin algebra to superalgebras.
In this case, both $r(\lambda)$ and $L(\lambda)$ have a
supermatrix structure however. Hence the tensor product in
(\ref{class.yb}) and (\ref{gaudin.compact}) must be
understood in a ${\bf Z}\sb 2$-graded sense. Precisely,
Eq.(\ref{gaudin.compact}) can be written as
\begin{eqnarray*}
&&L(\lambda)\otimes L(\mu) - (-1)\sp{\partial L(\lambda) \partial
L(\mu)}L(\mu)\otimes L(\lambda) \\
&&~~ = L(\lambda) r\sb{(1)}(\lambda-\mu)\otimes r\sb{(2)}(\lambda-\mu) +
(-1)\sp{\partial L(\mu) \partial
r\sb{(1)}}r\sb{(1)}(\lambda-\mu)\otimes L(\mu)r\sb{(2)}(\lambda-\mu)
\\
&&~~ - (-1)\sp{\partial L(\lambda) \partial
r\sb{(2)}}r\sb{(1)}(\lambda-\mu) L(\lambda)\otimes
r\sb{(2)}(\lambda-\mu) - r\sb{(1)}(\lambda-\mu)\otimes
r\sb{(2)}(\lambda-\mu)L(\mu),
\end{eqnarray*}
where we have used the notation $r(\lambda-\mu) =
r\sb{(1)}(\lambda-\mu)\otimes r\sb{(2)}(\lambda-\mu)$ and e.g.
$\partial L(\mu)$ denotes the parity of an appropriate element of
$L(\mu)$.

 In the present paper we describe  models which can be solved using
$osp(1,2)$ Gaudin algebra. In particular we generalise the Gaudin
model by  considering a system of $N$ hamiltonians
\begin{equation}
H\sb i = \sum\sb{j\neq i}{{\bf S}\sp i\cdot{\bf S}\sp j
+ V\sb +\sp i V\sb -\sp j - V\sb
-\sp iV\sb +\sp j\over{\epsilon\sb i - \epsilon\sb
j}}, \quad i= 1,\ldots ,N,
\label{osph}
\end{equation}
where the components of ${\bf S}\sp i$, satisfy the algebra
(\ref{su2}) and
\begin{eqnarray}
&&[S\sp i\sb z, V\sp j\sb \pm ]\sb - = \pm\h\delta\sb{ij}V\sp i\sb \pm
, \quad   [S\sp i\sb \pm , V\sp j\sb \pm ]\sb - = 0, \quad [S\sp i\sb
\pm , V\sp j\sb \mp ]\sb - = \delta\sb{ij}V\sp i\sb \pm \nonumber \\
&& [V\sp i\sb \pm , V\sp j\sb \pm ]\sb + = \mp\h\delta\sb{ij}S\sp i\sb
\pm  ,\quad [V\sp i\sb +, V\sp j\sb - ]\sb + = \mp\h\delta\sb{ij}S\sp
i\sb z,
\label{osp}
\end{eqnarray}
where $[\; ,\; ]\sb +$ denotes the anticommutator.

The paper is organised as follows. In Section~II we construct the
$osp(1,2)$ Gaudin algebra and we define a generating function for
integrals of motion of
 systems described by this algebra.
In Section~III we
construct the spectrum of the generating function and we derive the
Bethe equations.
In Section~IV we use the $osp(1,2)$ Gaudin algebra to solve the $\osp$
Gaudin model. We show that the hamiltonians (\ref{osph}) commute with
each other and construct explicitly their complete spectrum, treating
the two particle (N=2) special case as an example. In
section five  we discuss the $\osp$ Dicke model. This is a natural
$\osp$ generalisation of the ordinary Dicke model governed by an
$su(1,1)$ Gaudin algebra, a generalisation which is realised by
coupling a system of boson and fermion oscillators to the Gaudin model
itself. We
prove that the spectrum obtained for it by the algebraic Bethe Ansatz of
section three is complete. Then we focus our attention on the special
(N=1) case in which there is
only one particle in the underlying Gaudin model. In an oscillator
representation of
$\osp$, this case describes a system with two
oscillators with frequency ratio 2:1 together with one ordinary and one
Majorana fermionic oscillators.
We discuss the non-separability of the Schr\"odinger equation for this
case.
\vskip 1cm

\noindent {\bf II. STRUCTURE OF THE ${osp(1,2)}$ GAUDIN ALGEBRA}\vskip
.3cm

\noindent Here we define the $\osp$ Gaudin algebra. To do it we first
define a classical r-matrix and a quantum L-operator. The $r$-matrix
is constructed out of the quadratic Casimir of
$\osp$ in a standard way \cite{kulish3},
\begin{equation}
r(\lambda) = {{2}\over{\lambda}} (S\sb z\otimes S\sb z + \h(S\sb
+\otimes S\sb - + S\sb -\otimes S\sb +) + V\sb +\otimes V\sb - - V\sb
- \otimes V\sb +), \quad \lambda\in{\bf R},
\label{rmatrix.abs}
\end{equation}
written in the fundamental representation of $\osp$ (unitary in the
compact case, non-unitary in the non-compact case),
\begin{eqnarray}
&&S\sb z = \h\pmatrix{1&0&0\cr 0&0&0\cr 0&0&-1}, \quad  S\sb + = S\sb
-\sp t = \pmatrix{0&0&1 \cr 0&0&0\cr 0&0&0} \nonumber \\
&&V\sb + =\h\pmatrix{0&1&0\cr 0&0&1\cr 0&0&0} , \quad V\sb - =
\h\pmatrix{0&0&0\cr -1&0&0\cr 0&1&0}.
\label{fund}
\end{eqnarray}
Here and in what follows we use the convention, in which a matrix
element $a\sb{ij}$ is even (odd) if $i+j$ is even (odd). This
convention proves more suitable for our purposes, than the standard
one, in which a supermatrix is written in a canonical block form
(cf. \cite{dewitt1}). Inserting (\ref{fund}) into (\ref{rmatrix.abs})
we obtain an explicit expression for $r(\lambda)$
\begin{equation}
r(\lambda) = {1\over{2\lambda}}\pmatrix{1&0&0&&0&0&0&&0&0&0\cr
0&0&0&&-1&0&0&&0&0&0\cr 0&0&-1&&0&1&0&&2&0&0\cr\cr
0&1&0&&0&0&0&&0&0&0\cr
0&0&1&&0&0&0&&-1&0&0\cr 0&0&0&&0&0&0&&0&1&0\cr\cr
0&0&2&&0&-1&0&&-1&0&0\cr
0&0&0&&0&0&-1&&0&0&0\cr 0&0&0&&0&0&0&&0&0&1\cr}.
\label{rmatrix}
\end{equation}
Next we define the quantum $L$-operator
\begin{equation}
L(\lambda) \equiv  \pmatrix{A(\lambda) &V\sb-(\lambda)&B(\lambda)\cr
-V\sb+(\lambda)&0&V\sb -(\lambda)\cr C(\lambda)&V\sb+(\lambda)&
-A(\lambda)}.
\label{lmatrix}
\end{equation}
The components of $L(\lambda)$ generate the $osp(1,2)$ Gaudin algebra.
We assume that they
are non-singular except for a finite number of distinct values of
$\lambda$, namely $\epsilon \sb 1 , \ldots, \epsilon\sb N$, such that
$\epsilon\sb 1>\ldots >\epsilon\sb N$.

Using definitions (\ref{rmatrix}), (\ref{lmatrix}) and relation
(\ref{gaudin.compact}) we obtain the complete set of
relations defining the (rational) $osp(1,2)$ Gaudin algebra
\begin{eqnarray}
&&[A(\lambda),A(\mu)]\sb - = [B(\lambda),B(\mu)]_- =
[C(\lambda),C(\mu)]_- = 0, \nonumber \\
&&[B(\lambda),V_-(\mu)]_- =
[C(\lambda),V_+(\mu)]_ - = 0, \nonumber\\
&&[A(\lambda),C(\mu)]\sb - = -{{C(\lambda)-C(\mu)}\over{\lambda -\mu}},
\nonumber \\
&& [A(\lambda),B(\mu)]\sb - = {{B(\lambda)-B(\mu)}\over{\lambda
-\mu}}, \nonumber \\
&&[A(\lambda),V\sb \pm(\mu)]\sb - =
\mp\h{{V\sb\pm(\lambda)-V\sb\pm(\mu)}\over{\lambda -\mu}}, \nonumber \\
&&[B(\lambda),C(\mu)]\sb - = 2{{A(\lambda)-A(\mu)}\over{\lambda
-\mu}},\label{gaudin} \\
&& [V\sb-(\lambda),V\sb +(\mu)]\sb - =
\h{{A(\lambda)-A(\mu)}\over{\lambda -\mu}}, \nonumber \\
&&[B(\lambda),V\sb +(\mu)]\sb - =
{{V\sb-(\lambda)-V\sb-(\mu)}\over{\lambda -\mu}}, \nonumber\\
&&[C(\lambda),V\sb -(\mu)]\sb - =
{{V\sb+(\lambda)-V\sb+(\mu)}\over{\lambda -\mu}}, \nonumber \\
&&[V\sb -(\lambda), V\sb -(\mu)]\sb + = -\h{{B(\lambda)-B(\mu)}\over{\lambda
-\mu}}, \nonumber\\
&& [V\sb +(\lambda), V\sb +(\mu)]\sb + =
\h{{C(\lambda)-C(\mu)}\over{\lambda -\mu}}. \nonumber
\end{eqnarray}

In the $\osp$ Lie superalgebra there is a clear relation between
$V\sb\pm$ and $S\sb\pm$, i.e. $4V\sb\pm\sp 2 = \mp S\sb\pm$. Therefore
the universal enveloping algebra $U(\osp)$ is
generated by $V\sb\pm$ and $S\sb z$ only and the highest weight
representations are constructed by the actions of a single raising
operator $V\sb+$ say. In the Gaudin algebra the
similar relation can be read off the last two of Eqs. (\ref{gaudin}).
When $\mu$ approaches $\lambda$ we obtain
\begin{equation}
B'(\lambda) = 4 V\sb-^2(\lambda),\qquad
C'(\lambda) = -4V\sb+^2(\lambda) ,
\label{b-v}
\end{equation}
where prime denotes differentiation with respect to $\lambda$. We
learn from Eqs.
(\ref{b-v}) that construction of highest weight modules of $\osp$
Gaudin algebra involves both $V\sb +(\lambda)$ and $C(\lambda)$
operators.  This also makes
the Bethe Ansatz construction of the next section more complicated,
as compared to the $su(1,1)$ case, in which  there is only one raising
operator.

{}From the point of view of integrable models described by the
$osp(1,2)$ Gaudin algebra, it
is important to consider a generating function
\begin{eqnarray}
t(\lambda) & \equiv& \h\str L^2(\lambda) \nonumber \\
& = & A\sp 2(\lambda) +\h[B(\lambda),C(\lambda)]\sb + +
[V\sb+(\lambda), V\sb -(\lambda)]\sb -.
\label{t}
\end{eqnarray}
This definition implies
\begin{equation}
[t(\lambda),t(\mu)]\sb- = 0,
\label{commute}
\end{equation}
so that the function $t(\lambda)$ generates a family of commuting
hamiltonians. Solution  of an integrable model is equivalent to the
construction for it of the spectrum of $t(\lambda)$. This can be
carried out in the
framework of the algebraic Bethe Ansatz, as will be discussed in the
next section.\vskip 1cm

\noindent {\bf III. THE SPECTRUM OF $t(\lambda)$ }\vskip .3cm

\noindent Now we construct the spectrum of the generating function
$t(\lambda)$. We look for the highest weight representations of the
Gaudin algebra
(\ref{gaudin}) with a vacuum $|0\rangle$ defined by
\begin{equation}
V\sb-(\lambda)|0\rangle = B(\lambda)\vac = 0, \quad A(\lambda)\vac
=\alpha(\lambda)\vac .
\label{vacuum}
\end{equation}

Using definition (\ref{t}) of $t(\lambda)$ it is easy to find that
\begin{eqnarray}
t(\lambda)\vac & = & (\alpha^2(\lambda)+\h\alpha'(\lambda))\vac \nonumber
\\
& \equiv & \tau\sb 0 (\lambda)\vac ,
\label{tau0}
\end{eqnarray}
where the prime denotes differentiation with respect to $\lambda$.
Other eigenstates of $t(\lambda)$ are then generated from the vacuum $\vac$
by the repeated actions of $V\sb+(\lambda)$ and $C(\lambda)$.
Accordingly we look for an eigenstate  $|\mu\sb 1,
\ldots ,\mu_n\rangle$ of $t(\lambda)$  of the form
\begin{equation}
|\mu\sb 1, \ldots ,\mu_n\rangle = V\sb+(\mu\sb 1, \ldots ,\mu_n)\vac.
\label{eigenstate}
\end{equation}
Here $V\sb+(\mu\sb 1, \ldots ,\mu_n)$ is constructed from $V\sb+(\mu\sb
i)$ and $C(\mu\sb i)$, $i=1,\ldots ,n$. Demanding that $|\mu\sb 1,
\ldots ,\mu_n\rangle$ be an eigenstate of $t(\lambda)$ we determine the
form of $V\sb+(\mu\sb 1, \ldots ,\mu_n)$ and the numbers $\mu_1,\ldots
,\mu_n$. The corresponding eigenvalue of $t(\lambda)$ will be denoted
by $\tau(\lambda;\mu_1,\ldots ,\mu_n)$. The explicit construction of
$V\sb+(\mu\sb 1, \ldots ,\mu_n)$ is rather involved. We  list
operators $V\sb+(\mu\sb 1, \ldots ,\mu_n)$ for the first few
excitations
at the end of the section. Nevertheless it is possible
to find the eigenvalues $\tau(\lambda;\mu_1,\ldots ,\mu_n)$ as well as
numbers $\mu_1,\ldots ,\mu_n$ without performing a  fully explicit
construction of
$V\sb+(\mu\sb 1, \ldots ,\mu_n)$.

The operator $V\sb+(\mu\sb 1, \ldots ,\mu_n)$ may be decomposed as
follows
\begin{equation}
V\sb+(\mu\sb 1, \ldots ,\mu_n) = V\sb+(\mu\sb 1) \cdots V\sb+(\mu_n) +
W(\mu_1,\ldots ,\mu_n),
\label{v.decom}
\end{equation}
where $W(\mu_1,\ldots ,\mu_n)$ involves at least one operator
$C(\mu\sb i)$, i.e.
$$
W(\mu_1,\ldots ,\mu_n) = \sum_{\sigma\in
S_n \atop \sigma_1<\ldots <\sigma_{n-2}}\gamma_\sigma(\mu_1,\ldots
,\mu_n)V\sb+(\mu\sb{\sigma_ 1}) \cdots
V\sb+(\mu_{\sigma_{n-2}}) C(\mu_{\sigma_{n-1}}) + \ldots ,
%\label{decom}
$$
where $\gamma_\sigma(\mu_1,\ldots ,\mu_n)$ are number coefficients.
It is crucial to observe
that $W(\mu_1,\ldots ,\mu_n)$ cannot produce terms proportional to
$V\sb+(\mu\sb 1) \cdots V\sb+(\mu_n)$, when commuted with
$t(\lambda)$. Likewise $W(\mu_1,\ldots ,\mu_n)$ itself cannot generate
an eigenvector of $t(\lambda)$. Since (\ref{eigenstate}) defines an
eigenvector of
$t(\lambda)$, the eigenvalue $\tau(\lambda;\mu_1,\ldots ,\mu_n)$ has
to be equal to the coefficient of $V\sb+(\mu\sb 1) \cdots
V\sb+(\mu_n)$, when acted upon by $t(\lambda)$. We compute
\begin{eqnarray}
[t(\lambda),V\sb+(\mu\sb 1) \cdots V\sb+(\mu_n)]\sb - &=& V\sb+(\mu\sb
1) \cdots V\sb+(\mu_n) (\sum\sb{i=1}\sp n {1\over{\lambda-\mu\sb i}}
A(\lambda) +\h\sum\sb{i<j}{1\over{(\lambda-\mu\sb i)(\lambda-\mu\sb
j)}}) \nonumber \\
\nonumber\\
&& + \left({\rm terms\; involving\; at\; least \; one \; } C(\mu\sb
i)\right).
\end{eqnarray}
{}From this we immediately see that
\begin{equation}
\tau(\lambda;\mu_1,\ldots ,\mu_n) = \sum\sb{i=1}\sp n
{1\over{\lambda-\mu\sb i}} \alpha(\lambda)
+\h\sum\sb{i<j}{1\over{(\lambda-\mu\sb i)(\lambda-\mu\sb j)}} +
\tau\sb 0(\lambda).
\label{t.eigen}
\end{equation}

 The numbers $\mu\sb i$, $i=1,\ldots ,n$ are not yet specified. They can
be determined as follows \cite{kulish2}. We observe that the operator
$t(\lambda)$
is non-singular for any $\lambda\neq \epsilon\sb i$, $i=1,\ldots ,N$.
In particular, $\mu\sb i$, $i=1,\ldots ,n$ are regular points of
$t(\lambda)$. This implies that $\tau(\lambda;\mu_1,\ldots ,\mu_n)$
should be non-singular at $\lambda=\mu\sb i$, $i=1,\ldots ,n$.
The conditions for non-singularity in (\ref{t.eigen}) at
$\lambda=\mu\sb i$, $i=1,\ldots ,n$ read
\begin{equation}
\alpha(\mu\sb i) =-\h\sum\sb{j=1 \atop j\neq i}\sp{n} {1\over \mu\sb i
-\mu\sb j}, \qquad i=1,\ldots ,n.
\label{bethe}
\end{equation}
These are the Bethe equations for the $osp(1,2)$ Gaudin algebra.
Taking (\ref{bethe}) into
account we can write (\ref{t.eigen}) as
\begin{equation}
\tau(\lambda;\mu_1,\ldots ,\mu_n) = \tau\sb 0(\lambda) +
\sum\sb{i=1}\sp n{\alpha(\lambda)-\alpha(\mu_i)\over\lambda-\mu_i}.
\label{tau.newform}
\end{equation}

We can also write (\ref{t.eigen}) in the form \cite{sklyanin4}
\begin{equation}
\tau(\lambda;\mu_1,\ldots ,\mu_n) = (\chi(\lambda;\mu_1,\ldots ,\mu_n)
+ \alpha(\lambda))^2 +\h {d\over d\lambda}(\chi(\lambda;\mu_1,\ldots
,\mu_n) +\alpha(\lambda)),
\label{lame1}
\end{equation}
where
$$
\chi(\lambda;\mu_1,\ldots ,\mu_n) = \h {q_n'(\lambda)\over
q_n(\lambda)}, \quad q_n(\lambda) = \prod\sb{i=1}\sp n(\lambda-\mu_i).
$$
Eq.(\ref{lame1}), rearranged as a differential equation for
$q_n(\lambda)$, reads
\begin{equation}
q_n''(\lambda) +4\alpha(\lambda)q_n'(\lambda) + 4(\tau_0(\lambda) -
\tau(\lambda;\mu_1,\ldots ,\mu_n))q_n(\lambda) = 0,
\label{lame}
\end{equation}
so that (\ref{t.eigen}) and (\ref{bethe}) are equivalent to the
differential equation (\ref{lame}). This equation plays an important
role in the analysis of the completeness of the spectrum of
$t(\lambda)$ constructed by the Bethe Ansatz. We illustrate this
in the next section.

We would like to end this section with a few comments on the structure
of operators $V_+(\mu_1,\ldots ,\mu_n)$. Since each $\mu_i$
corresponds to a fermionic degree of freedom, the function
$V_+(\mu_1,\ldots ,\mu_n)$ is totally antisymmetric.  The first three
operators $V_+(\mu_1,\ldots ,\mu_n)$ come out as
\begin{eqnarray*}
&&V_+(\mu_1), \quad \underline{V_+(\mu_1)V_+(\mu_2)}
-\q{C(\mu_1)+C(\mu_2)\over \mu_1-\mu_2}, \\
&&\underline{V_+(\mu_1)V_+(\mu_2)V_+(\mu_3)} - {1\over 8}\sum\sb{i,j,k
= 1}\sp{3}\epsilon\sb{ijk}V_+(\mu_i){C(\mu_j)+C(\mu_k) \over
\mu_j-\mu_k}.
\end{eqnarray*}
Here $\underline{V_+(\mu_1)\cdots V_+(\mu_n)}$ denotes the antisymmetric
product
$$
\underline{V_+(\mu_1)\cdots V_+(\mu_n)} = {1\over n!}\sum\sb{\sigma\in
S\sb n}\sgn (\sigma) V_+(\mu_{\sigma(1)})\cdots V_+(\mu_{\sigma(n)}).
$$
It can be checked directly that these operators generate
eigenvectors of $t(\lambda)$, provided that conditions (\ref{bethe})
are satisfied.

Finally we notice that if there is no pseudovacuum (\ref{vacuum}),
but there is a pseudovacuum $V\sb+(\lambda)|0\rangle = C(\lambda)\vac =
0$, $A(\lambda)\vac =\tilde{\alpha}(\lambda)\vac$ instead, then the
whole of the construction  can be repeated with $V\sb+(\lambda)$
replaced by $-V\sb-(\lambda)$, $C(\lambda)$ replaced by $-B(\lambda)$
and $\alpha(\lambda)$ replaced by $-\tilde{\alpha}(\lambda)$.

\vskip 1cm

\noindent {\bf IV. THE $osp(1,2)$ GAUDIN MODEL}\vskip .3cm

\noindent
Here we discuss the model defined by the hamiltonians (\ref{osph}). We
show that  hamiltonians (\ref{osph}) are in involution and
 construct their common spectrum. We  also prove, this
spectrum is complete. Finally we  derive the spectrum
explicitly for the simplest non-trivial case of the two-particle
system.  We begin with a proof that the model is integrable.

We consider the $L$-operator
\begin{equation}
L(\lambda) =  \sum\sb{i=1}\sp N {1\over \lambda-\epsilon\sb
i}\pmatrix{S\sp i\sb z&V\sb -\sp
i&-S\sb-\sp i \cr -V\sb+\sp i & 0 & V\sb -\sp i\cr -S\sb+\sp i
&V\sb+\sp i &-S\sp i\sb z}.
\label{lmatrixg}
\end{equation}
Using the commutation rules (\ref{su2}) and (\ref{osp}), we can
directly check that $L(\lambda)$ generates the $osp(1,2)$ Gaudin
algebra. Writing
generators of the $osp(1,2)$ Gaudin algebra in the explicit form
(\ref{lmatrixg}), we easily
find that the generating function (\ref{t}) takes the following form
\begin{equation}
t(\lambda) = \sum\sb{i=1}\sp N {({\bf S}\sp i)\sp 2\over
(\lambda-\epsilon\sb i)\sp 2} +\sum\sb{i\neq j} {{{\bf S}\sp i\cdot{\bf
S}\sp j + V\sb +\sp i V\sb -\sp j - V\sb
-\sp iV\sb +\sp j}\over{(\lambda -\epsilon\sb i)(\lambda - \epsilon\sb
j)}}
\label{tgaudin}
\end{equation}
Hamiltonians (\ref{osph}) are identified with
\begin{equation}
H\sb i = \res\sb {\lambda = \epsilon\sb i} t(\lambda)
\label{h-t}
\end{equation}
and thanks to Eq.(\ref{commute}) they commute with each other.
They also satisfy $\sum\sb i H\sb i =0,$ so that
(\ref{h-t}) gives $N-1$ independent hamiltonians. It is obvious that
${\bf S}^2$ commutes with all $H\sb i$. Now it remains to show that
$S\sb z$ commutes with $t(\lambda)$
and hence with each of $H\sb i$'s. The easiest way to do it is first
to observe that the
algebra (\ref{gaudin}) will not change if we transform
$$
A(\lambda)\rightarrow A(\lambda) +g ,\qquad g\in {\bf R}.
$$
This transformation induces the transformation
$$
t(\lambda) \rightarrow t(\lambda) +g^2 + 2A(\lambda)
=\tilde{t}(\lambda).
$$
Obviously $\tilde{t}(\lambda)$ generates a family of commuting
hamiltonians and
$$
\tilde{H}\sb i = \res\sb{\lambda =\epsilon\sb i}\tilde{t}(\lambda) =
H\sb i +2gS^i_z.
$$
Now $\sum\sb i \tilde{H}\sb i =2gS\sb z$ and from $[\tilde{H}\sb
i,\tilde{H}\sb j]\sb -=0$ we get $[S\sb z, H\sb i]\sb -=0$ as claimed.

{}From now on we focus on the non-compact case, in which $\osp$ contains
$su(1,1)$ as a subalgebra. We show,
how to construct the complete spectrum of $t(\lambda)$
(\ref{tgaudin}). We define the numbers $s\sb i>0$, $i=1,\ldots ,N$ by
$$
S\sb z\sp i\vac = s\sb i\vac ,\qquad i=1,\ldots ,N.
$$
With these definitions
$$
\alpha(\lambda) = \sum\sb{i=1}\sp N {s\sb
i \over \lambda -\epsilon\sb i}$$
 and the Bethe equations (\ref{bethe})
take the form
$$
\sum\sb{i=1}\sp N {s\sb
i \over \lambda -\epsilon\sb i} = -\h\sum\sb{j=1 \atop j\neq i}\sp{n}
{1\over \mu\sb i
-\mu\sb j}, \qquad i=1,\ldots ,n.
$$
They can be identified with those corresponding to $su(1,1)$ Gaudin
model by the rescaling $s\sb i \to \h s\sb i$. Therefore,  as
in the $su(1,1)$ case, the differential equation
(\ref{lame}) becomes the Lam\'e equation, intensively studied in the
context of orthogonal polynomials \cite{szego1}. The result, which we
need here is the Heine-Stieltjes theorem \cite{szego1} asserting that
there are
$$
C\sb{N-1}\sp n ={n+ N-2\choose n}
$$
polynomial solutions to (\ref{lame}), corresponding to
$C\sb{N-1}\sp n$ different choices of $\tau(\lambda;\mu_1,\ldots
,\mu_n)$. Hence the Bethe Ansatz gives $C\sb{N-1}\sp n$ eigenstates of
$t(\lambda)$. On the other hand the Hilbert space of the system can be
identified with $\H\sb 1\otimes \cdots \otimes \H\sb N$, where $\H_i$,
$i=1,\ldots ,N$
is a Hilbert space of representation of $\osp$ corresponding to
the $i$-th particle. All eigenstates of $t(\lambda)$ are constructed
as linear combinations of basic states in $\H\sb 1\otimes \cdots
\otimes \H\sb N$, the latter obtained by the repeated actions of
$V\sb +\sp i$'s, $i=1,\ldots ,N$ on the vacua $\vac\sb i$. The number
of all possible eigenstates of $t(\lambda)$ at the $n$-th level equals
the number of all possible distributions of $n$ numbers between $N$
intervals and hence reads
$$
C\sb{N}\sp n ={n+ N-1\choose n}.
$$
This means that still $C\sb{N}\sp n - C\sb{N-1}\sp n =
C\sb{N}\sp{n-1}$ states need to be constructed. Using
an argument similar to the one that lead to $[t(\lambda), S\sb z]\sb -
=0$, we can
show that  $[t(\lambda), V\sb +]\sb - =0$, where $V\sb + =
\sum_{i=1}^NV\sb +\sp i$. The remaining states at the $n$-th level are
obtained by the action of $V\sb +$ on each state of  the $n-1$-th
level.

Since $V\sb +$ commutes with $t(\lambda)$, it is clear that at the
$n$-th level the full spectrum of $t(\lambda)$ consists of
$$
\tau_0(\lambda), \;\;\tau(\lambda ;\mu\sb 1\sp{(1)}),\;\; \ldots
,\;\;\tau(\lambda
;\mu_1^{(n)}, \ldots , \mu_n^{(n)}),
$$
corresponding to the states
$$
(V_+)^n\vac ,\;\;(V_+)^{n-1}V_+(\mu\sb 1\sp{(1)})\vac ,\;\; \ldots ,\;\;
V_+(\mu_1^{(n)}, \ldots , \mu_n^{(n)})\vac .
$$

To illustrate, how  the theory of orthogonal polynomials can be
used to solve the $\osp$ Gaudin model,  we look at the case
$N=2$, $\epsilon_1= -\epsilon_2 =\epsilon$, $\epsilon >0$,
$s_1=s_2 =s$. Without the loss of generality we can set $\epsilon =1$. We
construct the spectrum of $t(\lambda)$
explicitly for this simple model.
The Lam\'e equation (\ref{lame}) can be now written
\begin{equation}
(1-\lambda^2)q_n''(\lambda) -8s \lambda q_n'(\lambda) -
4(\tau_0(\lambda) - \tau(\lambda;\mu_1,\ldots
,\mu_n))(1-\lambda^2)q_n(\lambda) = 0, \label{jacobi.eq}
\end{equation}
where $q_n(\lambda)$ is a polynomial of degree $n$.
Eq.(\ref{jacobi.eq}) has a polynomial solution of degree $n$, provided
that
\begin{eqnarray}
\tau(\lambda;\mu_1,\ldots ,\mu_n) & = & \tau_0(\lambda) +{n(n+8 s
-1)\over 4(\lambda^2-1)} \nonumber \\
& = & \tau_0(\lambda) + {n(n+8 s
-1)\over 8 s\lambda}\alpha(\lambda).
\label{t.eigen.2}
\end{eqnarray}
If (\ref{t.eigen.2}) is satisfied, then
\begin{equation}
q_n(\lambda) = P_n^{(4 s -1,4 s -1)}(\lambda),
\label{jacobi}
\end{equation}
where $P_n^{(4 s -1,4 s -1)}(\lambda)$ denotes the Jacobi
polynomial. The numbers $\mu_i$, $i=1,\ldots ,n$ are then
zeros of the Jacobi polynomial (\ref{jacobi}). The full spectrum of
$t(\lambda)$ at the $n$-th excitation reads
$$
\{\tau_0(\lambda) + {k(k+8 s -1)\over
8 s\lambda}\alpha(\lambda);\; k=0,1,\ldots ,n\}.
$$

The $\osp$ algebra can be represented in a Fock space of the harmonic
oscillator as follows \cite{macfarlane4}
\begin{eqnarray}
&&S^i_z = {1\over 4}[a_i,\adag_i]_+, \quad S_-^i=-\h a_i^2, \quad
S_+^i = \h(\adag_i)^2 \nonumber \\
&&V_+^i = {c_i\adag_i\over \sqrt{8}}, \quad V_-^i =- {c_ia_i\over
\sqrt{8}}, \qquad i=1,2 \label{osp.oscillator} ,
\end{eqnarray}
where $a_i$, $\adag_i$ are destruction and creation operators,
$[a\sb i,\adag\sb j]\sb- =\delta\sb{ij}$ and $c_i^2 = 1$, $c_1c_2 =
-c_2c_1$. In this case $ s = {1\over 4}$ and the Jacobi polynomials
(\ref{jacobi}) become the Legendre polynomials $P_n(\lambda)$, and
the spectrum of $t(\lambda)$ reads
$$
\{ {k(k+1)(\lambda^2-1) -1\over
4(\lambda^2-1)^2};\; k=0,1,\ldots ,n\},
$$
so that the eigenvalues of hamiltonian $H\sb 1$ say, at the $n$-th
level are $k(k+1)/8$, $k=0,1,\ldots ,n$.

\vskip 1cm

\noindent {\bf V. THE DICKE MODEL}\vskip .3cm

\noindent
Here we present another model which satisfies the $osp(1,2)$ Gaudin
algebra and
hence can be solved in the way described in Section~III. This is the
model of a bosonic and a fermionic oscillator coupled to the $\osp$
Gaudin system. Let
$[b,\bdag]_-=1$ be a bosonic oscillator and let $[f,\fdag]_+=1$ be a
fermionic oscillator.
It is easy to check that the operators
\begin{eqnarray}
&&A(\lambda) = -\h \lambda + A^0(\lambda), \quad B(\lambda)
=b+B^0(\lambda), \quad C(\lambda) =-\bdag +C^0(\lambda) \nonumber\\
&&V_-(\lambda) =-\h f+V^0_-(\lambda), \qquad V_+(\lambda) =\h \fdag
+V^0_+(\lambda)
\label{dicke}
\end{eqnarray}
satisfy the algebra (\ref{gaudin}) provided that $A^0(\lambda)$,
$B^0(\lambda)$, $C^0(\lambda)$, $V^0_-(\lambda)$, $V^0_+(\lambda)$
satisfy it and $b,f$ oscillators (anti-)commute with them.
We assume that $A^0(\lambda)$, $B^0(\lambda)$, $C^0(\lambda)$,
$V^0_-(\lambda)$, $V^0_+(\lambda)$ correspond to the $N$-particle
$\osp$ Gaudin model and we show that the Bethe Ansatz in this case
gives all the eigenstates of $t(\lambda)$. In other words we show
that the Bethe equations (\ref{bethe}), which read explicitly in this
case
\begin{equation}
-\h\mu_i + \sum_{k=1}^ N{ s_k\over \mu_i-\epsilon_k} =
-\h\sum\sb{j=1 \atop j\neq i}\sp{n} {1\over \mu\sb i -\mu\sb j},
\qquad i=1,\ldots ,n,
\label{dicke.bethe}
\end{equation}
have precisely $C\sb{N+1}\sp n ={n+ N\choose n}$ solutions. Using the
classical method of Stieltjes (cf. \cite{szego1}) we consider the
function
\begin{equation}
Z(\mu_1,\ldots ,\mu_n) = e^{-\h\sum_{i=1}^n\mu_i^2}\prod_{k=1,\ldots,
N\atop j=1,\ldots
,n}|\mu_j-\epsilon_k|^{2 s_k}\prod_{r<s}|\mu_r-\mu_s|,
\label{action}
\end{equation}
which can be defined in $C\sb{N+1}\sp n$ regions in ${\bf R}^n$
corresponding to all
possible distributions of $n$ variables $\mu_i$ between $N+1$
intervals, $(-\infty ,\epsilon_n]$, $[\epsilon_n,\epsilon_{n-1}]$,
$\ldots$, $[\epsilon_1 ,+\infty)$. Let us fix one of these regions. It
is clear that at the boundary of this region $Z$ vanishes, and since
$Z\geq 0$, it has a maximum inside the region. At this maximum
$\mu_r\neq \mu_s$, if $r\neq s$, and ${\partial Z\over
\partial\mu_i}=0$, $i=1,\ldots ,n$, i.e. Eqs.(\ref{dicke.bethe}) hold.
Since there are $C\sb{N+1}\sp n$ regions, there are $C\sb{N+1}\sp n$
different maxima, hence (\ref{dicke.bethe}) has at least $C\sb{N+1}\sp
n$ real solutions. Knowing that there can be at most $C\sb{N+1}\sp n$
solutions to (\ref{dicke.bethe}) we have also proved that there are
precisely $C\sb{N+1}\sp n$ maxima of (\ref{action}).

Now we look at the simplest non-trivial case of the Dicke model. We
assume that $N=1$, $\epsilon\sb 1 =0$, and we write $\osp$ generators
in the oscillator representation (\ref{osp.oscillator}). In the purely
bosonic case this model corresponds to the system of two oscillators
with the frequency ratio $2:1$. In the $\osp$ case we have the model
of two such oscillators with, in addition, an ordinary and one
Majorana fermionic oscillators. The
generating function $t(\lambda)$ takes the form
\begin{equation}
t(\lambda) = {1\over 4} \lambda\sp 2 -\h H -\h G\lambda\sp{-1}
-{1\over 16}\lambda\sp{-2},
\label{t.oscillator}
\end{equation}
where
\begin{eqnarray}
H & = & \h[a,\adag]\sb + +[b,\bdag]\sb+ -\h[f,\fdag]\sb - ,\nonumber\\
G & = & b\adag\sp 2 +\bdag a\sp 2 -\rh fc\adag +\rh\fdag ca.
\label{gh.hamiltonians}
\end{eqnarray}

The use of (\ref{tau.newform}) gives immediately the eigenvalues of
$t(\lambda)$ in the form
\begin{equation}
\tau(\lambda ;\mu\sb 1 ,\ldots ,\mu\sb n) = {1\over 4}\lambda\sp 2
-{1\over 16}\lambda\sp{-2} - \h(n+1) -({1\over 4}\sum\sb i\sp
n{1\over{\mu\sb i}})\lambda\sp {-1}.
\end{equation}
Therefore the hamiltonians $H$ and $G$ have the eigenvalues
\begin{equation}
E\sb n = n+1 ,\qquad g\sb n =\h \sum\sb i\sp n{1\over{\mu\sb i}}
\label{eg}
\end{equation}
respectively. The eigenvalues $g\sb n$ need more discussion. From the
symmetries of the function $Z(\mu_1,\ldots ,\mu_n)$ (\ref{action}), we
immediately deduce that if $\mu\sb i$ is a solution to
(\ref{dicke.bethe}) so is $-\mu\sb i$, hence the spectrum of $G$ is
closed under the change of sign. In addition it contains 0 for $n$
even. So the spectrum of $G$ is fully characterised by the positive
eigenvalues. On the other hand $g\sb n$ can be derived directly from
the differential equation (\ref{lame}), which now takes the form
\begin{equation}
\lambda q''\sb n(\lambda) - (2\lambda^2 - 1)q'\sb n(\lambda) +
2(n\lambda + g\sb n)q\sb n(\lambda) = 0.
\label{2osc.lame}
\end{equation}
Here we have used the explicit form (\ref{eg}) of the eigenvalue $E\sb n$
of $H$. Making the ansatz, $q\sb n(\lambda) = \sum\sb{k=0}\sp n a\sb
k\lambda\sp k$, $a\sb n =1$ we can derive the recurrence relations for
$a\sb k$,
\begin{eqnarray}
&&a\sb{n-1} =-g\sb n \nonumber \\
&&(n-k)\sp 2 a\sb{n-k} + 2(k+2)a\sb{n-k-2} +2g\sb na\sb{n-k-1} =0, \quad
k=0,1,\ldots ,n-1.
\label{an.recur}
\end{eqnarray}
Solving (\ref{an.recur}) for $n=1,2$, we find that the
eigenvalues of $G$ are $\pm\sqrt{\h}$ and $0,\pm\sqrt{3}$ respectively.

We can represent bosonic oscillators $a$, $b$ in terms of the canonical
coordinates, $x\sb i$, $p\sb i$, $i=1,2$, $[x\sb i,p\sb j]\sb -
=i\delta\sb{ij}$, as follows
\begin{equation}
a=\rh (x\sb 2 + ip\sb 2) ,\qquad b=x\sb 1 +{i\over 2}p\sb 1.
\end{equation}
We can also represent fermions $f, c$ in terms of the Pauli matrices
$\sigma\sb i$,
\begin{equation}
c=\sigma\sb 3 ,\qquad f=\h(\sigma\sb 1-i\sigma\sb 2).
\end{equation}
In this representation, the hamiltonians (\ref{gh.hamiltonians}) take
the form
\begin{eqnarray}
H & = &\h\pmatrix{p\sb 1\sp 2 + p\sb 2\sp 2 +4x\sb 1\sp 2 +x\sb 2\sp 2+ 1
& 0 \cr\cr 0 & p\sb 1\sp 2 + p\sb 2\sp 2 +4x\sb 1\sp 2 +x\sb 2\sp 2- 1},
\nonumber \\
% && \nonumber\\
\label{hg.cart}\\
G & = & \h\pmatrix{-2x\sb 1p\sb 2 \sp 2 +p\sb 1[x\sb 2, p\sb 2]\sb+
+2x\sb 1 x\sb 2\sp 2 & -x\sb 2 -ip\sb 2 \cr\cr -x\sb 2 + ip\sb 2 &-2x\sb
1p\sb 2 \sp 2 +p\sb 1[x\sb 2, p\sb 2]\sb+ +2x\sb 1 x\sb 2\sp 2}.\nonumber
\end{eqnarray}

It is known \cite{boyer1} that, in the purely bosonic case, Schr\"odinger
equations corresponding to $G$ and $H$ can be separated in parabolic
coordinates $y\sb 1$, $y\sb 2$,
\begin{equation}
2x\sb 1 = y\sb 1\sp 2- y\sb 2\sp 2 ,\qquad x\sb 2 = y\sb 1y\sb 2.
\label{parabolic}
\end{equation}
It is also known that these are the only coordinates other than
Cartesian's, in which the bosonic system separates. We can try
to implement this procedure in the $\osp$ case. First we write
hamiltonians (\ref{hg.cart}) in parabolic coordinates
(\ref{parabolic}),
\begin{eqnarray}
H & = & {1\over 2(y\sb 1\sp 2 +y\sb 2\sp 2)}\pmatrix{-{\partial\sp
2\over \partial y\sb 1\sp 2} +y\sb 1\sp 6 +y\sb 1\sp 2 -{\partial\sp
2\over \partial y\sb 2\sp 2} +y\sb 2\sp 6 +y\sb 2\sp 2 & 0 \cr\cr
0 & -{\partial\sp
2\over \partial y\sb 1\sp 2} +y\sb 1\sp 6 -y\sb 1\sp 2 -{\partial\sp
2\over \partial y\sb 2\sp 2} +y\sb 2\sp 6 -y\sb 2\sp 2 } \nonumber \\
\label{hg.para}\\
G & = & {y\sb 1\sp 2 y\sb 2\sp 2 \over 2(y\sb 1\sp 2 + y\sb 2\sp
2)}\pmatrix{{1\over y\sb 2\sp 2}{\partial\sp
2\over \partial y\sb 2\sp 2} -y\sb 2\sp 4 -{1\over y\sb 1\sp 2}{\partial\sp
2\over \partial y\sb 1\sp 2} + y\sb 1\sp 4 & -{1\over y\sb 1y\sb
2}({1\over y\sb 1}{\partial\over \partial y\sb 1} + y\sb 1\sp 2 +
{1\over y\sb 2}{\partial\over \partial y\sb 2} + y\sb 2\sp 2) \cr\cr
{1\over y\sb 1y\sb
2}({1\over y\sb 1}{\partial\over \partial y\sb 1} - y\sb 1\sp 2 +
{1\over y\sb 2}{\partial\over \partial y\sb 2} - y\sb 2\sp 2) &
{1\over y\sb 2\sp 2}{\partial\sp
2\over \partial y\sb 2\sp 2} -y\sb 2\sp 4 -{1\over y\sb 1\sp 2}{\partial\sp
2\over \partial y\sb 1\sp 2} + y\sb 1\sp 4}.\nonumber
\end{eqnarray}

We look for vector wave functions
\[ \Psi\sb n(y\sb 1,y\sb 2) = \pmatrix{\Psi\sb n\sp +(y\sb 1,y\sb 2)
\cr \Psi\sb n\sp -(y\sb 1,y\sb 2)}, \]
which solve the Schr\"odinger equations
\begin{equation}
H\Psi\sb n = E\sb n\Psi\sb n ,\quad G\Psi\sb n = g\sb n\Psi\sb n.
\label{schrodinger}
\end{equation}
It can be easily seen that the first of Eqs.(\ref{schrodinger}) can be
solved, assuming that $\Psi\sb n(y\sb 1,y\sb 2)$ separates, i.e.
\begin{equation}
\Psi\sb n(y\sb 1,y\sb 2) = \pmatrix{\Phi\sb n\sp +(y\sb 1)
\tilde{\Phi}\sb n\sp +(y\sb 2)
\cr \Phi\sb n\sp -(y\sb 1) \tilde{\Phi}\sb n\sp -(y\sb 2)}.
\label{separate}
\end{equation}
In this case the first of Eqs.(\ref{schrodinger}) leads to the four
1-dimensional equations
\begin{eqnarray}
&& {-\Phi\sb n\sp\pm}'' (y\sb 1\sp 6 -(2n+2\pm 1)y\sb 1\sp 2
-c\sb\pm)\Phi\sb n\sp\pm =0 ,\nonumber \\
&&{-\tilde{\Phi}\sb n\sp\pm}'' (y\sb 2\sp 6 -(2n+2\pm 1)y\sb 2\sp 2
+c\sb\pm)\tilde{\Phi}\sb n\sp\pm =0,
\label{separate2}
\end{eqnarray}
where $c\sb\pm$ are constants. The equations for
$\Phi\sb n\sp -$ are the same as for $\Phi\sb{n+1}\sp +$, hence we
need to consider only equations for $\Phi\sb{n+1}\sp +$ say.
Eqs.(\ref{separate2}) are precisely identical with the equations  one
obtains in the purely bosonic case. From this we immediately know,
that all independent solutions to the first of Eqs.(\ref{schrodinger})
can have the separated form (\ref{separate}). Inserting
(\ref{separate}) into the second of Eqs.(\ref{schrodinger}) we
immediately see that if $\Phi\sb n\sp\pm$, $\tilde{\Phi}\sb n\sp\pm$
satisfy (\ref{separate2}), they cannot satisfy the Schr\"odinger
equation for $G$. One has to combine different solutions of
(\ref{separate2}) to obtain the function $\Psi\sb n(y\sb 1,y\sb 2)$.
Knowing that the parabolic coordinates (\ref{parabolic}) are the only
ones, for which separation of variables in the first of
Eqs.(\ref{schrodinger}) is possible, we immediately deduce that we
cannot separate (\ref{schrodinger}) changing only the position
variables.

One can nevertheless expect that the separation of
(\ref{schrodinger}) is possible, but a more general canonical
transformation of the Cartesian variables has to be considered. An
indication comes from the analysis of the bosonic case. In that case,
the zeros of corresponding functions $\Phi\sb n$ can be identified with
the zeros of the polynomial $q\sb n$, satisfying the bosonic version of
Eq.(\ref{2osc.lame}). It is therefore reasonable  to hope that
there are coordinates, in which Schr\"odinger equations
(\ref{schrodinger}) give such a natural meaning to
Eq.(\ref{2osc.lame}).

\vskip 1cm

\noindent {\bf VI. ACKNOWLEDGEMENTS}\vskip .3cm

\noindent T. Brzezi\'nski would like to thank St. John's College,
Cambridge for a
Benefactors' Scholarship. His work is also supported by the grant KBN
2 0218 91 01.

\newpage

\baselineskip 17pt
% \bibliographystyle{unsrt}
% \bibliography{ref}

\begin{thebibliography}{10}

\bibitem{gaudin1}
M.~Gaudin.
\newblock Diagonalisation d'une classe d'hamiltoniens de spin.
\newblock {\em J. de Physique}, 37:1086--1098, 1976.

\bibitem{gaudin2}
M.~Gaudin.
\newblock {\em La fonction d'onde de Bethe}.
\newblock Masson, Paris, 1983.

\bibitem{jurco4}
B.~Jur\v co.
\newblock Classical {Yang}-{Baxter} equations and quantum integrable
systems.
\newblock {\em J. Math. Phys}, 30:1289, 1989.

\bibitem{sklyanin4}
E.K. Sklyanin.
\newblock Separation of variables in the {Gaudin} model.
\newblock {\em J. Sov. Math.}, 47:2473--2488, 1989.

\bibitem{neumann1}
C.~Neumann.
\newblock De problemate quodam mechanico, quod ad primam integralium
  ultraellipticorum classem revocatur.
\newblock {\em Reine und Angew. Math.}, 56:46--63, 1859.

\bibitem{kuznetsov1}
V.B. Kuznetsov.
\newblock Quadrics on real {Riemmanian} spaces of constant curvature:
  {Separation} of variables and connection with {Gaudin} magnet.
\newblock {\em J. Math. Phys.}, 33:3240--3254, 1992.

\bibitem{jurco3}
B.~Jur\v co.
\newblock On quantum integrable models related to nonlinear quantum optics.
  {An} algebraic {Bethe} {Ansatz} approach.
\newblock {\em J. Math. Phys}, 30:1739, 1989.

\bibitem{belavin2}
A.A. Belavin and V.G. Drinfeld.
\newblock Triangle equations and simple {L}ie algebras.
\newblock {\em Sov. Sci. Rev. C}, 4:93--165, 1984.

\bibitem{kulish3}
P.P. Kulish.
\newblock Quantum {O}{S}{P}-invariant nonlinear {Schr{\"o}dinger} equation.
\newblock {\em Lett. Math. Phys.}, pages 87--93, 1985.

\bibitem{dewitt1}
B.~DeWitt.
\newblock {\em Supermanifolds}.
\newblock CUP, Cambridge, 1992.

\bibitem{kulish2}
P.P. Kulish and E.K. Sklyanin.
\newblock Quantum spectral transform method. {Recent} developments.
\newblock In J.~Hietarinta and C.~Montonen, editors, {\em Lecture Notes in
  Physics 151. Integrable Quantum Field Theories, Tvarminne, Finland 1981},
  pages 61--119. Springer-Verlag, 1982.

\bibitem{szego1}
G.~Szeg{\"o}.
\newblock {\em Orthogonal Polynomials}.
\newblock AMS, New York, 1939.

\bibitem{macfarlane4}
A.J. Macfarlane and S.~Majid.
\newblock Quantum group structure in a fermionic extension of the quantum
  harmonic oscillator.
\newblock {\em Phys. Lett.}, B268:71--74, 1991.

\bibitem{boyer1}
C.P. Boyer and K.B. Wolf.
\newblock The 2:1 anisotropic oscillator, separation of variables and
symmetry  group in {Bargmann} space.
\newblock {\em J. Math. Phys.}, 16:2215--2223, 1975.

\end{thebibliography}

\end{document}